\documentclass[pre,twocolumn,showpacs,amsmath,amssymb]{revtex4}

\usepackage{dcolumn}
\usepackage{bm}

 \usepackage{times}

\usepackage{balance} 

\usepackage{graphicx} 
\usepackage{lastpage}
\usepackage[format=plain,justification=raggedright,singlelinecheck=false,font=small,labelfont=bf,labelsep=space]{caption} 
\usepackage{fancyhdr}
\usepackage{rotating}
\usepackage{multirow}

\newcommand{\ur}{{\boldsymbol{\hat{u}_r}}}

\newcommand{\utheta}{\boldsymbol{\hat{u}_\theta}}

\newcommand{\mfric}{{\zeta}_R}

\newcommand{\fric}{{\zeta_R}^\bot}
\newcommand{\MSAD}{\langle \Delta \psi^2(t) \rangle}
\newcommand{\Dr}{D_{R}}
\newcommand{\Lm}{L}
\newcommand{\T}{\mathrm{T}}
\newcommand{\dg}{\,^\circ}

\begin{document}

\title{Rotational microrheology of Maxwell fluids using micron-sized wires}

\author{R\'emy Colin$^{*}$\footnotetext{$^*$Present address: Rowland Institute at Harvard, 100 Edwin H Land Bvd, Cambridge, MA 02141, USA}}
\author{Loudjy Chevry}
\author{Jean-Fran\c cois Berret} 
\author{B\'ereng\`ere Abou}

\affiliation{Laboratoire
  Mati\`ere et Syst\`emes Complexes (MSC), UMR CNRS 7057 \& Universit\'e
  Paris Diderot, FRANCE} 

\begin{abstract}
We demonstrate a simple method for
  rotational microrheology in complex fluids, using micrometric wires. The three-dimensional
  rotational Brownian motion of the wires suspended in Maxwell fluids is measured from their
  projection on the focal plane of a microscope. We analyze the mean-squared angular displacement
  of the wires of length between $1$ and $ 40\ \mu$m. The
  viscoelastic properties of the suspending fluids are extracted from
  this analysis and found to be in good agreement with macrorheology
  data. Viscosities of simple and complex fluids between $10^{-2}$ and $30$
Pa.s could be measured. As for the elastic modulus, values up to $\sim 5$ Pa could be determined. This simple technique, allowing for a broad range of probed length scales, opens new perspectives in microrheology of heterogeneous materials such as gels, glasses
  and cells.
 \end{abstract}

\maketitle

\section{introduction}

Microrheology consists in using microscopic probe particles embedded within a fluid to
measure the relation between stress and deformation\cite{Breeveld,Review-TWaigh-2005,AnRevFlu-Mason-2010,cicuta}. In
passive microrheology, the linear viscoelastic properties of the fluid are
derived from the thermal motion of the
probes. In contrast, active
microrheology involves forcing probes externally, and can be extended to the nonlinear regime of deformation\cite{BiophysJ-Ziemann-1994,PRE-Cappallo-2007,dhar2010}. 

Microrheology experiments can be performed on small volumes, typically
one microliter, which is essential when a limited amount of material
is available, e.g. in biological samples. The technique was recently
pushed to the limit, with less than one nanolitre available to investigate the
beetle secretion \cite{JRSI-Abou-2010}. Microrheology has opened up new fields of
investigation in soft materials and complex fluids, including cells\cite{hoffman_consensus_2006,weihs_bio-microrheology:_2006,
  mackintosh_active_2010}. By overcoming the
limitations of traditional bulk rheology, it gives access to heterogeneities and an extended
range of probed frequency and moduli\cite{Review-TWaigh-2005}. Lastly, the technique has been used in
out-of-equilibrium glasses, by combining passive and active methods to
investigate deviations to fluctuation-dissipation
relations\cite{Abou-PRL-2004,PRL-Jabbari-2007}.

Most microrheology experiments are based on the investigation of the
translational thermal motion of isotropic probes, typically spherical latex or
silica colloids. In the case of anisotropic probes, the rotational
motion becomes accessible as well. Despite recent developments in the
synthesis of nano- and micrometric anisotropic particles
\cite{xia_one-dimensional_2003,
  tang_one-dimensional_2005,srivastava_nanoparticle_2009,Frka-Petesic2011},
rotational microrheology has rarely been investigated. This comes mainly from the difficulty of
recording the rotational thermal motion in
three-dimension (3D) which requires advanced optical
techniques\cite{Science-Han-2006,Chang2010,Xiao2011}. For instance,
rotational Brownian motion was studied using light streak tracking of
thin microdisks~\cite{PRL-Cheng-2003}, depolarized
dynamic light scattering and epifluorescence microscopy of optically
anisotropic spherical colloidal
probes~\cite{PRL-Andablo-2005,Langmuir-Anthony-2006}. It was also investigated with scanning
confocal microscopy of colloidal rods with 3D resolution~\cite{JCIS-Mukhija-2007}, or reconstruction of the wire
position from its hologram observed on the focal plane of a
microscope~\cite{OE-Cheong-2010}. Recently, we have proposed a method to measure the out-of-plane rotational motion of rigid wires, simply using 2D video microscopy \cite{colin-rods-2012}. 

Anisotropic probes such as rigid wires present advantages over
spherical probes. The translational Brownian motion of spherical
probes up to $5 \,\mu$m can typically be detected with standard video
microscopy techniques. By comparison, we have previously shown that
the translational and rotational diffusion of rigid wires up to $100 \,\mu$m could be measured\cite{colin-rods-2012}. With rigid
wires, the viscoelastic properties of materials such as actin
networks \cite{PRL-Amblard-1996, PRE-Lu-2002} or
cells\cite{tseng_micromechanical_2002,hoffman_consensus_2006}, can be
probed in a wide range of length scales, typically $1-100$ microns, giving
a complete picture of the material. This is of importance in complex fluids which typically
exhibit hierarchical structures on separate length scales. 

In this paper, we demonstrate the ability of a recent technique for
measuring the rotational thermal motion to probe the viscoelastic
properties of complex fluids. The model complex fluids we used are
wormlike micelle solutions, which are well characterized Maxwell
fluids\cite{Lerouge2010} and have been already investigated with other
microrheology
techniques\cite{PRE-Wilhelm-2003,Buchanan2005,Willenbacher2007}. Wires
between $1$ and $40\,\mu$m in length are immersed in the Maxwell
fluids with elastic moduli between $1-80$ Pa, and viscosities between
$0.1-35 $ Pa.s. The 3D rotational Brownian motion of the wires is
extracted from their 2D projection on the focal plane of a microscope,
following the procedure developed in \citet{colin-rods-2012}. The
viscoelastic parameters of the Maxwell fluids, elastic modulus and
static viscosity, are then deduced, and found in good agreement
with the rheological measurements. The resolution of the method is
shown to depend on the wire length.

\section{Rotational thermal motion of rigid wires: theory}
\label{sec:theory}
The rotational Brownian motion of a rigid micrometric wire in a viscous fluid can be modeled by a Langevin equation describing the
fluctuations of the wire orientation unit vector $\ur$\footnote{$(r, \theta, \varphi)$ is the spherical
  coordinates system for a wire diffusing in a three-dimensional
  space.}. In the
absence of an external torque, and neglecting inertia, the rotational
equation of motion reads~\cite{Doi-Edwards}:
\begin{equation}
\boldsymbol{0} = -\mfric \boldsymbol{\omega} + \boldsymbol{\Gamma_r}
\label{rotationEquation}
\end{equation}
in which $\mfric \boldsymbol{\omega}$ and ${\boldsymbol{\Gamma_r}}$ are
respectively the viscous drag and the random Langevin torque. The
projection of Eq. (\ref{rotationEquation}) along the vector $\utheta$,
leads to :
\begin{equation}
\label{3DLangevin}
\fric \sin\theta\ \dot\varphi = \Gamma_1(t)
\end{equation}
where $\Gamma_1(t)$ is the component of the random torque on vector
$\utheta$ and $\fric$ the friction coefficient perpendicular to the wire
axis. The spherical coordinates $\varphi(t)$ and $\theta(t) $
describe the wire orientation\cite{colin-rods-2012}, and  $\dot\varphi = d\varphi/dt$. 
The variable $\psi(t)$,
defined as $\dot\psi = d\psi/dt=\sin\theta\ \dot\varphi$, obeys a one-dimensional Langevin equation (\ref{3DLangevin}) from which the mean-squared angular displacement (MSAD) is deduced~:
\begin{equation}
\label{eqn-psi}
\langle \Delta\psi^2(t) \rangle = 2 \frac{k_B\mathrm{T}}{\fric} t= 2 D_{R} t
\end{equation}
with $D_{R}$ the wire rotational diffusion coefficient and $k_B T$ the thermal energy. 

For a cylindrical wire, length $L$ and
diameter $d$, the perpendicular friction coefficient $\fric$ can be expressed as :
\begin{equation}
\fric = \frac{\pi \eta L^3}{3 \,g\left( L/d \right)}
\label{frictionDef}
\end{equation}
where $g\left( L/d \right)$ is a dimensionless function of the aspect ratio $p=L/d$ and $\eta$ is the fluid viscosity. According to \citet{JCP-Tirado-1984}, we assume here that $g(p)= \ln(p)
-0.662+0.917 \,p -0.050 \,p^2$, valid for $2<p<20$. The rotational diffusion coefficient $D_R$, deduced from Eqs. (\ref{eqn-psi}) and
(\ref{frictionDef}), writes:
\begin{equation}
\Dr = \frac{3k_B\mathrm{T}}{\pi\eta \Lm^3} g(\Lm / d)
\label{DiffCoefTheory}
\end{equation}

In a viscoelastic
fluid, Eq. (\ref{3DLangevin}) is extended in a generalized Langevin equation :
\begin{equation}
\int_{-\infty}^\infty dt'\zeta_R^\bot (t-t') \dot\psi(t')=\Gamma_1(t)
\end{equation}
where $\zeta_R^\bot (t)$ is a delayed friction kernel which takes into
account the fluid viscoelastic properties. The fluid is considered to be a linear viscoelastic material, where the coupling
between translation and rotation is a negligible second order
effect\cite{squires2010}. Considering that the surrounding stationary
medium is in thermal equilibrium at temperature $T$, a Generalized
Rotational Einstein Relation can be derived by analogy with the
translational case \cite{Abou-Physica-2008}:
\begin{equation}
\label{GRER}
s^2 \langle  \hat{\Delta\psi^2}(s) \rangle=2 k_B T/ \hat{\zeta}_R^\bot (s)
\end{equation}
In Eq. (\ref{GRER}), $\langle \hat{\Delta\psi^2}(s)
\rangle=\int_0^\infty \langle \Delta\psi^2(t) \rangle
\mathrm{e}^{-st} \,\mathrm{d}t$ and $\hat{\zeta_R}^\bot (s)$ are the Laplace transform of the MSAD and the friction coefficient respectively. Assuming that
Eq. (\ref{frictionDef}) remains valid in the viscoelastic material,
and can be extended to all frequencies $s$, the bulk frequency
dependent viscosity $\hat{\eta}(s)$ is related to the friction coefficient according to :
\begin{equation}
\hat{\zeta_R}^\bot (s)= \frac{\pi \hat{\eta}(s) L^3}{3 \,g\left( L/d \right)}
\label{frictionDef-s}
\end{equation}
From Eqs. \ref{GRER} and  \ref{frictionDef-s}, the relation between the complex modulus $\hat{G}(s)=s \hat{\eta}(s)$ and the MSAD is~:
\begin{equation}
s \langle  \hat{\Delta\psi^2}(s) \rangle=\frac{6k_BT \,g\left( L/d \right)}{\pi L^3 \hat{G}(s)}
\end{equation}

In a Maxwell fluid, the complex modulus writes $\hat{G}(s)=\eta_0 /(\tau_R + 1/s)$, where $\eta_0=G_0 \tau_R$ is the static viscosity, $\tau_R$ the
relaxation time and $G_0$ the plateau modulus. This yields the expression of the MSAD in a Maxwell fluid :
\begin{equation}
\label{max}
\langle \Delta\psi^2(t)\rangle=\frac{6k_BT \,g\left( L/d \right)}{\pi L^3 \eta_0}(t+\tau_R)
\end{equation}

The viscoelastic properties of the sample depend on the observation time scale. For $t \ll \tau_R$, the fluid behaves as an elastic solid
and the MSAD approaches a constant limiting plateau value \cite{van_zanten_brownian_2000}:
 
\begin{equation}
\label{max0}
\langle \Delta\psi^2(t \to 0)\rangle=\frac{6k_BT \,g\left( L/d \right)}{\pi L^3 G_0}
\end{equation}
For  $t \gg \tau_R$, the fluid behaves as a viscous liquid and the mean-squared angular
displacement increases linearly with time, with a slope proportional to $g(L/d)/L^3 \eta_0$. 

\section{Materials and Methods}
\label{sec:matmeth}


\subsection{Magnetic wires synthesis and characterization}
The wires were formed by electrostatic complexation between oppositely
charged nanoparticles and copolymers
\cite{Fresnais-AdMat-2008,yan_magnetic_2011}. The particles were $8.3$
nm iron oxide nanocrystals ($\gamma-\mathrm{Fe}_2\mathrm{O}_3$,
maghemite) synthesized by polycondensation of metallic salts in
alkaline aqueous media \cite{massart_preparation_1995}. To improve the
colloidal stability, the cationic particles were coated with
$M_W=2100\ \mathrm{g.mol}^{-1}$ poly(sodium acrylate) (Aldrich) using
the precipitation-redispersion process \cite{berret_size_2007}. This
process resulted in the adsorption of a resilient $3$ nm polymer layer
surrounding the particles. The copolymer used for the wire synthesis
was poly(trimethylammoniumethylacrylate)-b-poly(acrylamide) with
molecular weights $11000\ \mathrm{g.mol}^{-1}$ for the charged block
and $30000\ \mathrm{g.mol}^{-1}$ for the neutral block
\cite{berret-macromol-2007}. The applied protocol consisted first in
the screening of the electrostatic interactions by bringing the
polymer and particle dispersions to high salt concentration. In a
second stage, the salt was progressively removed by dialysis. To
stimulate their unidirectional growth, the particles and the polymers
were co-assembled in the presence of a magnetic field of $0.1$
Tesla. The shelf life of the co-assembled structures is of the order
of years. The wires are polydisperse. Their length distribution is
described by a log-normal function with median length $L_0=27$ $\mu$m
and polydispersity $s_{NW}=0.65$ \footnote{Throughout the manuscript,
  the polydispersity is defined as the ratio between the standard
  deviation and the average value.}. For this sample, the wires length
varies between $1$ and $40$ $\mu$m. The average diameter $d$ of the wires is
estimated at $400$ nm with scanning electron microscopy\cite{colin-rods-2012}. Electrophoretic mobility
and $\zeta-$potential measurements made with a Zeta sizer Nano ZS
Malvern Instrument showed that the wires are electrically neutral
\cite{yan_magnetic_2011}.

\subsection{Maxwell fluids}

 Wormlike micellar solutions have received considerable attention
 during the past three decades because of their remarkable rheological
 properties \cite{Lerouge2010}. The surfactant solutions investigated
 here are binary mixtures made of cetylpyridinium chloride
 (CP$^+$; Cl$^-$) and sodium salicylate (Na$^+$; Sal$^-$) (abbreviated as
 CPCl/NaSal) dispersed in a $0.5$ M NaCl brine
 \cite{Berret1994, Walker1996}. Since the pioneering work of Rehage
 and Hoffman, CPCl/NaSal is known to
 self-assemble spontaneously into micrometer long wormlike
 micelles\cite{rehage_rheological_1988}. Three CPCl/NaSal samples at concentration $c = 1$ wt. $\%$, $c = 2$ wt. $\%$ and $c = 6$ wt. $\%$ were investigated in this work. These wormlike micelles build a semi-dilute entangled
 network that imparts to the solution a Maxwell viscoelastic
 behavior. In the semi-dilute regime, the mesh size of the network decreases as $c^{-0.65}$, in good agreement with theoretical expectations\cite{Cates-1987}. It was found to be $35$ nm at $c = 1$ wt. $\%$, and $15$ nm at $c = 5$ wt. $\%$, i.e. much smaller than the wires smallest dimension\cite{Berret-langmuir-1993}. 

\subsection{Linear macrorheology}

The complex viscoelastic modulus $G^* (\omega)=G' (\omega)+i G''
(\omega)$ of the wormlike micellar solutions was measured with a
controlled shear rate rheometer (Physica MCR 500, cone and plate
geometry). The measurements were carried out for angular frequencies
in the range $\omega=0.1-100$ $\mathrm{rad.s}^{-1}$ in the linear
regime (temperatures in the range $\T=20-25\dg$C). A deformation of
$10\%$ was applied to the samples at $c = 2$ wt. $\%$ and $c = 6$ wt. $\%$,
while a $20\%$ deformation was applied to the sample at $c = 1$ wt. $\%$. The
viscoelastic response of CPCl/NaSal wormlike micelles was the one of a
Maxwell fluid with a unique relaxation time $\tau_R$. The viscoelastic
parameters $G_0$, $\tau_R$ and $\eta_0=\mathrm{lim}_{\omega\rightarrow
  0⁡}|G^* (\omega)|/\omega$ were derived from the measurements.

\subsection{Rotational particle tracking microrheology} 
\begin{figure}
\center
\includegraphics[width=8.0cm]{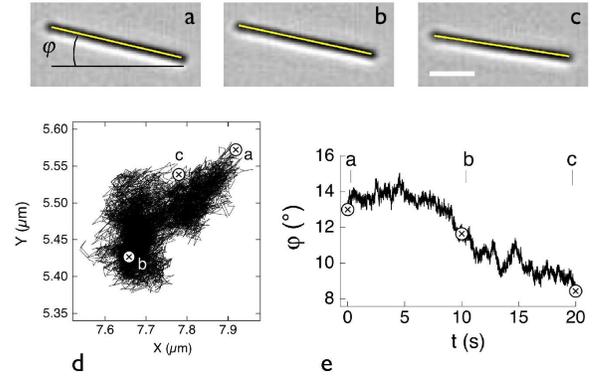}
\caption{(a,b,c) Images of a $9.2 \,\mu$m long wire immersed in a CPCl/NaSal wormlike micellar solution at $c=1$ wt. $\%$, at time intervals $0, 10$ and $20$ s. (d) Brownian motion of the wire center-of-mass in the objective focal plane $(x,y)$. (e) The time dependence of the angle $\varphi(t)$, defined with
respect to the horizontal axis, is extracted from the 2D images. The white scale bar is $5 \, \mu$m long. }
\label{fluc}
\end{figure}
\begin{figure*}[ht]
\centering
\includegraphics[width=16.0cm]{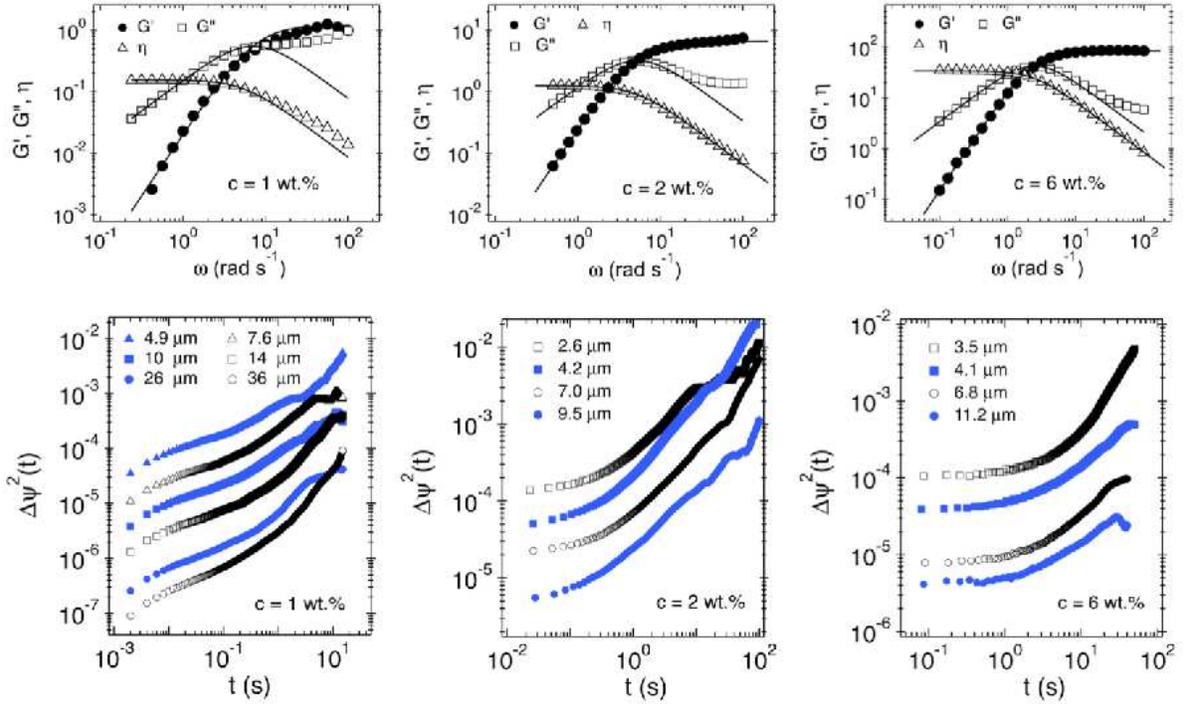}
\caption{Macrorheology (top) and microrheology (bottom) experiments in
  the CPCl/NaSal micellar solutions at $c = 1$ wt.$ \%$ ($T = 20
  \dg$C), $c = 2$ wt.$ \%$ ($T = 25 \dg$C) and $c = 6$ wt.$ \%$ ($T =  25 \dg$C). Top: elastic modulus $G'(\omega)$, loss modulus
  $G''(\omega)$, and magnitude of the complex viscosity
  $|\eta^*(\omega)|$. The solid lines correspond to the Maxwell
  model. Bottom: mean-squared angular displacement of wires immersed
  in the micellar solutions, as a function of the lag time. At short
  times, the MSADs exhibit a smooth increase (in the $c = 1$ wt. $\%$)
  or a plateau (in the $c = 2$ wt. $\%$ and $c = 6$ wt. $\%$). This is
  followed by an increase at longer times in all cases. These two
  regimes are identified as Regime I and Regime II, respectively. The
  transition from Regime I to Regime II occurs at a lag time which is
  solution dependent. For the $c = 1$ wt. $\%$, this lag time
  corresponds to the relaxation time $\tau_{R}$, whereas for the $c =
  2$ wt. $\%$ and $c = 6$ wt. $\%$, it is the time where the
  MSAD reaches the resolution limit (see section \ref{sec:result}).}
\label{macro-micro}
\end{figure*}

The 2D projection of the wires thermal fluctuations on the focal plane
of a microscope objective was recorded with a fast camera (EoSens
Mikrotron). The camera is coupled to an inverted microscope (Leica DM IRB) with a $100\,\times$ oil immersion
objective (NA$ = 1.3$). The objective temperature is controlled within $0.1\dg$C using
a Bioptechs heating ring coupled to a home-made cooling device. The sample temperature is controlled through the immersion
oil in contact. 
The camera was typically recording $100$ images per second during
$100$ s. The tracking of the wires
was made at least $60\,\mu$m from the observation chamber walls, to minimize the interactions between walls and wires. To
reduce the effect of hydrodynamic coupling, the wires concentration
was chosen around $c = 0.01$ wt. $\%$. Sedimentation of the wires was
negligible on the recording time scales.

Fig. \ref{fluc}-a,b,c display a $9.2 \,\mu$m long wire immersed in a CPCl/NaSal wormlike micellar solution at $c=1$ wt. $\%$, at time intervals $0, 10$ and $20$ s. The wire
orientation is determined by the angle $\varphi(t)$ defined with
respect to the horizontal axis, and shown in Fig. \ref{fluc}-e. The Brownian motion of the wire center-of-mass is shown in Fig. \ref{fluc}-d. The 3D Brownian motion
of the wire is extracted from its 2D projection on the
$(x,y)-$plane, according the procedure described in
\citet{colin-rods-2012}. The angle $\varphi(t)$ and the apparent
length $L_{app} (t)$ are measured from the 2D images, using a homemade
tracking algorithm implemented as an ImageJ plugin. The quantity
$\Delta\psi(t) = \psi (t) - \psi (0) = \int_0^t d t'
\sin\theta(t')\ \dot\varphi(t')$, and subsequently the MSAD,
$\langle\Delta\psi^2(t) \rangle$, reflecting the out-of-plane Brownian
motion of the wires, are then computed from $\varphi(t)$ and
$L_{app}(t)$. In \citet{colin-rods-2012}, the variable $\psi(t)$ was shown to correctly describe the 3D Brownian motion of the
wires in a viscous liquid.

\section{Results and discussion}
\label{sec:result}

\subsection{Macrorheology of wormlike micelles}
\label{macro}
Macrorheology experiments were performed on CPCl/NaSal
solutions at concentration $c = 1$ wt. $\%$, $c = 2$ wt. $\%$ and $c
= 6$ wt. $\%$. At these concentrations, the micellar solutions are known to exhibit a Maxwell viscoelastic behavior\cite{rehage_rheological_1988,Berret1994}, in agreement with the predictions of the Cates model\cite{Cates-1987}. 
Fig. \ref{macro-micro} (top) shows the elastic and loss moduli,
$G'(\omega)$ and $G''(\omega)$, and the magnitude of the complex
viscosity, $|\eta^{*}(\omega)|$, for the different concentrations. The data were adjusted with the expressions corresponding to
Maxwell fluids (solid lines),
$G'(\omega)/G_0=\omega^2\tau_R^2/(1+\omega^2\tau_R^2)$, 
$G''(\omega)/G_0=\omega\tau_R/(1+\omega^2\tau_R^2)$, and $|\eta^{*}(\omega)|=\eta_0/\sqrt{1+\omega^2\tau_R^2}$, with
$\eta_0=G_0\tau_R$ the static viscosity. 

The viscoelastic
parameters, $G_0$, $\tau_R$ and $\eta_0$, obtained from the adjustments are shown in Table
\ref{g-eta-tau}. They are in good agreement with macrorheology results published two
decades ago\cite{Berret1994,Lerouge2010}. As seen in Fig. \ref{macro-micro} (top), the linear viscoelastic responses
of the $c = 2$ wt. $\%$ and $c = 6$ wt. $\%$ wormlike micelles are found to be 
purely that of Maxwell fluids
\cite{Berret1994,Walker1996,Lerouge2010}, in the investigated
frequency range. By comparison, in the $c = 1$ wt. $\%$, slight deviations between
the data and the Maxwell model are observed at high frequency, arising from additional
relaxation mechanisms, such as the Rouse and breathing motions of the
micellar chains \cite{Graneck1992}.
\begin{table}[h]
\small
  \caption{Viscoelastic Maxwell parameters, $G_0$, $\tau_R$ and $\eta_0$, of the CPCl/NaSal micellar solutions (concentration $c$ in wt.), determined from macrorheology
    experiments.}
  \label{g-eta-tau}
  \begin{tabular*}{0.5\textwidth}{@{\extracolsep{\fill}}llll}
    \hline
    Sample & $G_0$ (Pa) & $\tau_R$ (s) & $\eta_0$ (Pa.s)\\
    \hline
     $c=1\%$, $T=20 \dg$C & 1.2 $\pm$ 0.2 & 0.12 $\pm$ 0.02 & 0.14 $\pm$ 0.05 \\
     $c=2\%$, $T=25 \dg$C & 8.0 $\pm$ 0.3 & 0.21 $\pm$ 0.01& 1.68 $\pm$  0.05\\
     $c=6\%$, $T=25 \dg$C & 83 $\pm$ 1 & 0.42 $\pm$ 0.02& 35 $\pm$ 1 \\
    \hline
  \end{tabular*}
\end{table}
\subsection{Rotational thermal motion of the wires}
\label{section:macromicro}
Microrheology experiments were carried out in the same micellar
solutions. Fig. \ref{macro-micro} (bottom) shows the MSADs of wires
immersed in the micellar solutions, as a function of the lag time. At
short times, the MSADs exhibit a slight increase (in the $c = 1$
wt. $\%$) or a plateau (in the $c = 2$ wt. $\%$ and $c = 6$ wt. $\%$). This is
followed by an increase at longer times in all cases. In the
following, these two regimes will be identified as Regime I and Regime
II, respectively. Before further analyzing
the data, the angular resolution of the
wire-based microrheology technique will be discussed.

\subsection{Angular resolution}
\label{angularRes}
\begin{figure}
\centering
\includegraphics[width=8.5cm]{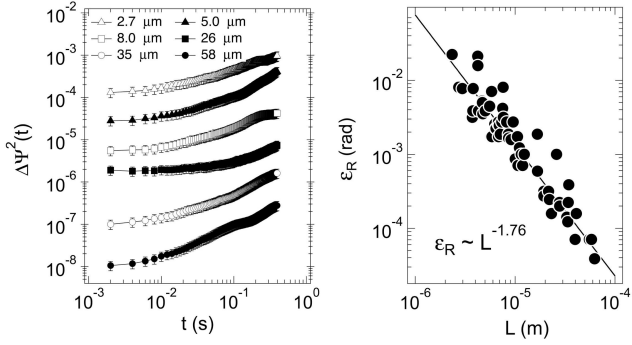}
\caption{Angular resolution of the wire-based microrheology
  technique. Left: MSAD measured in a viscous liquid of high viscosity
  ($92$ vol.$\%$ glycerol/water solution, $\eta=0.45$ Pa.s, $T=25 \dg$C) for
  wires between $2$ and $58\,\mu$m, in the same experimental
  conditions as in the viscoelastic fluids. Due to the angular
  resolution limit, the MSAD exhibits a plateau at short displayed lag
  times. Right: angular resolution $\varepsilon_{R}$ as a function of
  the wire length.  }
\label{epsilon}
\end{figure}
The angular resolution was determined from the wires fluctuations in
a viscous liquid of high viscosity $\eta$ ($92$ vol. $\%$
glycerol/water solution, $\eta=0.45$ Pa.s, $T=25\dg$C). According to Eq. \ref{eqn-psi}, the MSAD is expected to increase linearly with time. Fig. \ref{epsilon} (left) shows the MSAD for wires between $2$ and
$58\,\mu$m. At short lag time, the MSAD exhibits a plateau due to the
angular resolution limit, meaning that the minimum MSAD measurable with our setup is reached. Choosing a highly viscous fluid
ensures that this limit will be reached within an accessible time. 

Following the procedure developed for spherical probes
\cite{Savin2005}, the expression of the MSAD is extended as\cite{colin-rods-2012}~:
\begin{equation}
\langle \Delta\psi^2(t,L)\rangle=\frac{6k_BT \,g\left( L/d \right)}{\pi L^3 \eta}(t-\sigma/3) +2\,\varepsilon_{R}^2(L)\, ,
\end{equation}
which now includes the static error, $2 \varepsilon_{R}^2(L)$, and
the dynamic error accounting for the finite camera exposure time
$\sigma$. The extrapolation $\langle \Delta\psi^2(t \to 0,L)\rangle= 2 \varepsilon_{R}^2(L)
-2/3 D_{R}(L)\,\sigma$ provides the minimum detectable mean-squared
angular displacement. Since here $D_{R} \sigma/3 \ll
\varepsilon_{R}^2(L)$ on the entire range of $L$, this minimum reduces to $ 2 \varepsilon_{R}^2(L)$. 

Fig. \ref{epsilon} (right) shows $\varepsilon_{R}(L)$ as a function of
the wire length $L$. It decreases with increasing length, as $2 \times
10^{-12} L^{-1.76}$, with $\varepsilon_{R}$ in radian and $L$ in
meter. This expression gives an idea of the angular resolution,
$\sqrt{2} \varepsilon_{R}$, of our technique. It decreases from
$1.3\dg$ for $2\, \mu$m wires to $0.02\dg$ for $20\, \mu$m wires. This
yields a maximum measurable elastic modulus which
depends on the wire length, as expressed in Eq. \ref{max0} when the
MSAD reaches the resolution limit, $2 \varepsilon_{R}^2(L)$. To give an order of magnitude, it
corresponds to $2$ Pa for $2 \,\mu$m wires and $12$ Pa for $20\, \mu$m
wires. Longer wires will be more appropriate to study materials
of high elastic moduli.

Knowing our technique limitation in $G_0$, and based on the macrorheology results described
in section \ref{macro}, only the elastic modulus $G_0$ of the $c = 1$
wt. $\%$ CPCl/NaSal micellar solution can be determined with the
wire-based microrheology technique. For the $c = 2$
wt. $\%$ and $c = 6$
wt. $\%$, the elastic modulus $G_0$ are beyond the resolution limit, Regime I reflects the angular resolution limit, and only Regime II will be analyzed.

\subsection{Microrheology of wormlike micelles }
\label{analysis}
\begin{figure}
\center
\includegraphics[width=5.5cm]{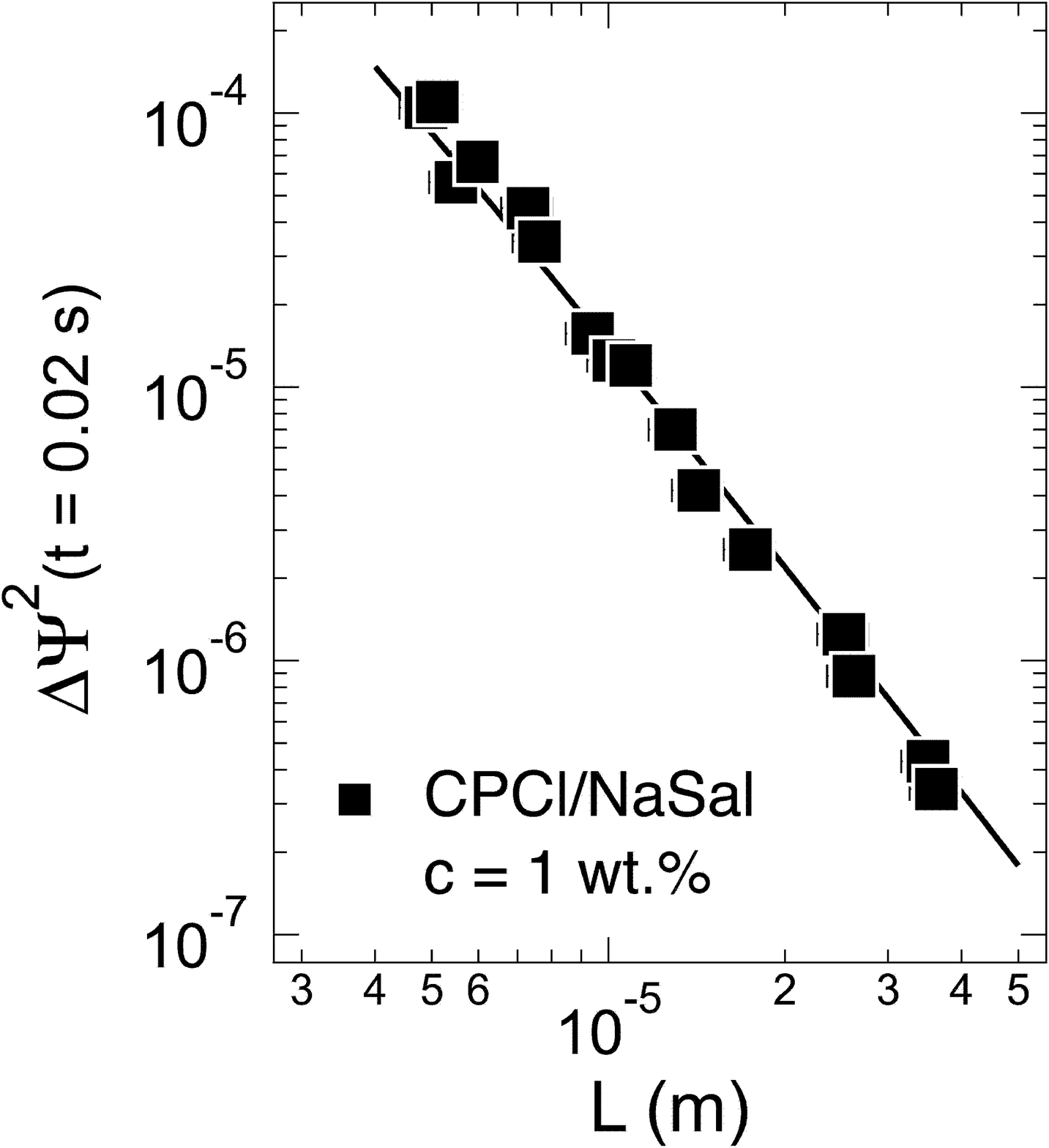}
\caption{Microrheology in the $c=1$ wt $\%$ CPCl/NaSal micellar solution ($T=20\dg$C). The mean-squared angular displacement in Regime I, $\langle \Delta\psi^2(t= 0.02 \,\rm{s}) \rangle$,
   is plotted as a function of the wire length $L$. At leading order, it decreases as
  $L^{-3}$. The elastic plateau modulus $G_0=1.5 \pm 0.5$ Pa is obtained by adjusting
the data with Eq. \ref{max0}, derived from the Maxwell model (solid line). It is in good agreement with the one measured with macrorheology, as shown in Table \ref{g-eta-tau}.}
\label{Drot}
\end{figure}

We now turn to the analysis of the MSAD measured in the micellar
solutions (Fig. \ref{macro-micro} (bottom)). 

In Regime I of the $c=1$ wt $\%$ CpCl/NaSal
solution, the MSADs extrapolated at $t \to 0$ were estimated as a
function of the wire length. Fig. \ref{Drot} shows a strong decrease of the data with
increasing $L$. Using Eq. \ref{max0} to
adjust the data in the form $\langle \Delta\psi^2(t \to 0) \rangle \sim g(L/d)L^ {-3}/ G_0$ (solid
line), the elastic plateau
modulus $G_0$ is deduced\footnote{At short time, in the plateau region, the MSAD can be
  modeled by a power-law behaviour $\MSAD \sim t^a$, with
  $a=0.3$. Given a power law behavior, the MSAD was corrected
  according to \citet{Savin2005} to take into account the static and dynamic errors.}. Assuming an
average diameter $d=400$ nm, least-square calculations provide
$G_0=1.5 \pm0.5$ Pa compatible with macrorheology. Note that in the $c =
1$ wt. $\%$ solution, the rather smooth plateau is the consequence of the extra Rouse and
breathing modes previously mentioned \cite{Graneck1992}. In the $c=2$ wt
$\%$ and $c=6$ wt $\%$ solutions, Regime I reflects the angular
resolution limit, as shown in section \ref{angularRes}. 

\begin{figure}
\center
\includegraphics[width=6.0cm]{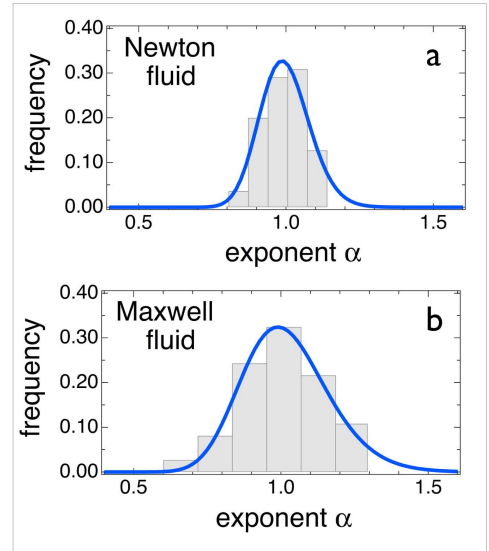}
\caption{Exponent distribution: the MSADs were adjusted using a power
law of the form $\langle \Delta\psi^2(t)\rangle \sim t^\alpha$,
$\alpha$ being an adjustable parameter. The distribution of exponents $\alpha$ is shown in a) for a Newtonian fluid (water/glycerol mixture) in b) for the micellar solutions. In
both cases, the exponent distribution is peaked around $1$, with a
similar polydispersity. These results evidence the existence of a
diffusive regime in the rotational fluctuations of wires immersed in
the micellar solutions (Regime II). }
\label{exponent}
\end{figure}

At longer times (Regime II), an increase of the MSADs versus time is
observed at all concentrations of the micellar solutions. The MSADs were adjusted using a power
law of the form $\langle \Delta\psi^2(t)\rangle \sim t^\alpha$,
$\alpha$ being an adjustable parameter. The distribution of the exponents $\alpha$ for the CPCl/NaSal
solutions is shown in Figure \ref{exponent} and compared to those
found for a Newtonian fluid ($50$ vol. $\%$
glycerol/water mixture) where
$\alpha$-values close to $1$ are expected. In
both cases, the exponent distribution is peaked around $1$, with a
similar polydispersity. These results evidence the existence of a
diffusive regime in the rotational fluctuations of wires immersed in
the micellar solutions.

\begin{figure}
\center
\includegraphics[width=7.0cm]{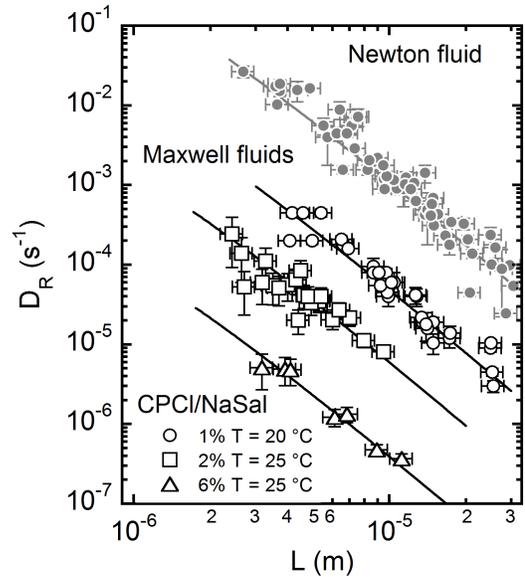}
\caption{Rotational diffusion coefficient versus length, for wires
  dispersed in a water/glycerol mixture ($50$ vol.$\%$, $ T=20\dg$C)
  (labeled “Newton fluid”) and in the CPCl/NaSal micellar solutions
  (labeled “Maxwell fluids”). The data for $D_{R}$ in the micellar
  solutions were adjusted according to Eq. (\ref{DiffCoefTheory})
  (solid lines), with the static viscosity $\eta_0$ as a parameter ($d
  = 400$ nm). The obtained values for $\eta_0$ are in good agreement
  with the macrorheology data, as shown in Table \ref{eta}.}
\label{allDrot}
\end{figure}

Based on the exponent analysis, the MSAD in Regime II is adjusted
according to $\langle \Delta\psi^2(t) \rangle = 2 D_{R} t$ extracted
from Eq. \ref{max}. The rotational diffusion coefficient $D_{R}$ in
the micellar solutions is shown in Fig. \ref{allDrot}, as a function
of the wire length. The results obtained in a water/glycerol mixture
($50$ vol. $\%$, $ T=20\dg$C) are also displayed. In all four fluids,
the rotational diffusion coefficient $D_{R}$ is found to decrease with
increasing length $L$. Using Eq. \ref{DiffCoefTheory} to adjust the
data in the form $D_{R}=3k_{B}T g(L/d)L^ {-3}/ \pi\eta_0$ (solid
lines), the static viscosity $\eta_0$ is extracted (Table
\ref{eta}). They are found to be in good agreement with the
macroscopic rheology data, validating our wire-based microrheology
technique. 

In Fig. \ref{macro-micro}, the transition between Regime I and Regime II
occurs at a lag time which is solution dependent. For the $c = 1$
wt. $\%$, this lag time corresponds to the relaxation time
$\tau_{R}$ (see Table \ref{g-eta-tau}), whereas for the $c = 2$ wt. $\%$ and $c = 6$ wt. $\%$, it
is the time where the MSAD reaches the
resolution limit.

\begin{table}
\caption{Static viscosities of the CPCl/NaSal solutions, and a
  glycerol/water mixture (($50$ vol.$\%$, $T=20\dg$C) obtained from
  macrorheology (tabulated value for the glycerol/water mixture) and
  wire-based microrheology. The data obtained with both techniques are
  found to be in good agreement. The values are given in format mean
  $\pm$ standard deviation.}  \center \setlength{\tabcolsep}{8pt}
\renewcommand{\arraystretch}{1.7}
\begin{tabular}{cccc}
\hline
\multicolumn{2}{c}{\multirow{2}{*}{samples}} & \multicolumn{2}{c}{$\eta_0$ (Pa.s)}\\ \cline{3-4}
\multicolumn{2}{c}{}  & rheometer & wires \\
\hline
\multicolumn{2}{c}{glycerol/water ($50$ vol.$\%$)} &  $0.0084$ & $0.0090 \pm 0.0003$ \\
\multirow{3}{*}{\begin{sideways}
CPCl/NaSal
\end{sideways}} & $c=1\,\%\ \T=20\dg$C & $0.14 \pm 0.05$ & $0.20 \pm 0.08$ \\ 
& $c=2\,\%\ \T=25\dg$C & $1.68 \pm 0.05$ & $1.7 \pm 0.5$ \\ 
& $c=6\,\%\ \T=25\dg$C & $35 \pm 1$ & $26 \pm 6$ \\ \hline
\end{tabular}
\label{eta}
\end{table}

Fig. \ref{allDrot} illustrates that $D_{R}(L)$ can be measured over at
least {\em $3$ decades} yielding the same range in accessible
viscosities. In our investigation, complex fluids with viscosities up
to $\sim 30$ Pa.s could be measured. The dispersion of the data for
$D_{R}(L)$, observed with respect to the fitting curves, is the
consequence of the wire diameter distribution~\footnote{The
  distribution in diameter is intrinsic to the wire fabrication
  method, and characterized here by a polydispersity of $0.4$.}, as
already seen in Newtonian fluids in
\citet{colin-rods-2012}. 

\section{Conclusion}

In this paper, we demonstrate the ability of a wire-based rotational
microrheology technique to probe the viscoelastic properties of
complex fluids. Wires of $400$ nm in diameter and $1-40 \,\mu$m in
length were synthesized by electrostatic complexation. Passive
microrheology was performed in CPCl/NaSal wormlike micellar solutions
with Maxwellian behavior. The out-of-plane rotational Brownian motion of the wires
was extracted from the 2D video microscopy images, following a method developed in \citet{colin-rods-2012}. The mean-squared
angular displacement of wires between $1-40\,\mu$m was measured in the Maxwell
fluids. Two regimes were identified. At short time, the MSAD exhibits
a plateau or pseudo-plateau, which is followed by an increase at
longer times. The first regime was shown to reveal the elastic plateau
modulus in the Maxwell fluids with low elasticity such as the $c = 1$
wt.$ \%$ CPCl/NaSal micellar solution. In the $c = 2$ wt.$ \%$ and $c
= 6$ wt.$ \%$ Maxwell fluids with higher elastic plateau modulus, it
reflects the angular resolution limit. In the second regime, the MSAD
was found to increase linearly, and was associated with a diffusive
behavior. The values of the static viscosities could be deduced. All
the values deduced from microrheology experiments, $G_0$ and $\eta_0$,
were found to be in good agreement with macrorheology measurements. Elastic components up to $\sim 5$ Pa could be measured, higher elasticities falling beyond the resolution
limit of the technique.
Viscosities in the range $10^{-2}$ to $30$
Pa.s were also measured. All the rheological properties were found to be independent of the wire length. 

Finally, we show here that micron-sized wires are efficient to probe
the viscoelastic properties of complex fluids, and allow for the
determination of both elastic and viscous components. With rigid
wires, the viscoelastic properties of materials can be probed in a
wide length scale range, typically $1-100\,\mu$m\cite{colin-rods-2012}, giving a complete picture of the
material. This is of importance in complex fluids which exhibit
hierarchical structures on separate length scales, and could exhibit
lengthscale-dependent rheology. This method broadens the array of tools available to microrheologists
studying complex fluids. The range of measurable viscoelastic
properties is similar to the one available with translational
particle-tracking microrheology \cite{Review-TWaigh-2005}. The main
advantage of this technique is to increase by two orders of magnitude
the range of accessible length scales, with a small number of probes involved compared to two-point microrheology\cite{PRL-Crocker-2000}.

\bibliography{biblio-all-rod} 
\bibliographystyle{apsrev} 

\end{document}